\pdfoutput=1

\documentclass[aip,jcp,reprint,groupedaddress,footinbib]{revtex4-1}

\usepackage[version=3]{mhchem}
\usepackage{graphicx} 
\usepackage{threeparttable}

\usepackage{amsmath} %

\usepackage{braket} %
\usepackage{dcolumn} %
\newcolumntype{d}[1]{D{.}{.}{#1}} %
\newcommand{\head}[1]{\multicolumn{1}{c}{#1}}
\newcommand{\HCOOH}{\ce{(HCOOH)2}}
\newcommand{\DCOOH}{\ce{(DCOOH)2}}
\newcommand{\HCOOD}{\ce{(HCOOD)2}}
\newcommand{\DCOOD}{\ce{(DCOOD)2}}

\newcommand{\double}[2]{\begin{tabular}{@{}c@{}}#1\\#2\end{tabular}}
\newcommand{\iu}{\ensuremath{\mathrm{i}}}
\newcommand{\eu}{\mathrm{e}^}

\newcommand{\fig}[2][]{Fig.~\ref{fig:#2}#1}

\newcommand{\rmd}{\mathrm{d}}

\newcommand{\Ref}[1]{Ref.~\onlinecite{#1}}

\newcommand{\tref}[1]{Table \ref{tab:#1}}
\newcommand{\wn}{\ensuremath{\text{cm}^{-1}}}
\usepackage{url}
\usepackage{hyperref}
\hypersetup{colorlinks=true,linkcolor=black,citecolor=black,urlcolor=black} %

\begin{document}

\title{Full- and reduced-dimensionality instanton calculations of the tunnelling splitting in the formic acid dimer}

\author{Jeremy O. Richardson}
\email{jeremy.richardson@phys.chem.ethz.ch}
\affiliation{Laboratory of Physical Chemistry, ETH Zurich, 8093 Zurich, Switzerland.}

\date{\today}

\begin{abstract}
The ring-polymer instanton approach is applied to compute the ground-state tunnelling splitting of four isotopomers of the formic acid dimer 
using the accurate PES of Qu and Bowman [\textit{Phys. Chem. Chem. Phys.}, 2016, \textbf{18}, 24835].
As well as performing the calculations in full dimensionality,
we apply a reduced-dimensionality approach to study how the results converge as successively more degrees of freedom are included.
The instanton approximation compares well to exact quantum results where they are available
but shows that nearly all the modes are required to quantitatively obtain the tunnelling splitting.
The full-dimensional instanton calculation reproduces
the experimental results,
with an error of only about 20 percent.
\end{abstract}

\maketitle

\section{Introduction}

The formic acid dimer forms a cyclic complex with a double hydrogen bond
and has been extensively studied as an example of this important non-covalent interaction. 
One of the best probes of the intermolecular forces \cite{StoneBook} is
through the tunnelling splitting 
caused by a concerted double proton transfer.
\cite{Birer2009review,HavenithHandbook}

A series of experimental studies by Havenith and coworkers \cite{Birer2009review,HavenithHandbook}
report measurements for the tunnelling splitting of various isotopomers of the formic acid dimer.
Although the initial assignment for the ground-state tunnelling splitting in \DCOOH\ was ambiguous,\cite{Madeja2002DCOOH}
it was later assigned as 0.0125\,\wn\ 
by comparison with the 0.016\,\wn\ splitting found in \HCOOH, \cite{Ortlieb2007HCOOH}
which is also in agreement with more recent experiments. \cite{Goroya2014HCOOH}
The splittings for \HCOOD\ and \DCOOD\ have not been observed
although an upper limit of 0.002\,\wn\ has been reported for the latter.
\cite{Gutberlet2008DCOOD}

Computing the tunnelling splitting of
the formic acid dimer
is a difficult challenge for theoretical chemistry. \cite{Birer2009review}
Due to the double proton transfer,
the barrier is significantly higher than in other
molecular systems such as malonaldehyde with only one hydrogen bond
and thus the tunnelling splitting is smaller by a few orders of magnitude.
As is often the case in theoretical chemistry,
there are two difficulties that must be overcome,
which are obtaining an accurate description of the potential-energy surface (PES)
and an accurate simulation of the quantum dynamics in many degrees of freedom.

The first of these challenges has recently been overcome by
Qu and Bowman, \cite{Qu2016formic} 
who developed
an PES
for the formic acid dimer
fitted to thousands of ab initio energies.
Although the PES appears to be very accurate,
they were nevertheless unable to reproduce the observed splittings
using an exact quantum dynamics calculation on a reduced-dimensionality system of up to four modes.
Note that, due to the small splitting, the statistical errors of a full-dimensional
diffusion Monte Carlo (DMC) calculation
have rendered this approach unusable. \cite{Qu2016formic}

In this paper,
we employ the ring-polymer instanton approach,
which has been used to describe tunnelling in a number of molecular systems.
\cite{Andersson2009Hmethane,rpinst,tunnel,water,octamer,hexamerprism}
It is derived from a well-understood approximation 
to quantum statistics
which is known to be increasingly accurate for high barriers with small splittings. \cite{tunnel}
It therefore complements the DMC method which is applicable to smaller barriers with large splittings.
The approach is very efficient and scales well with the number of degrees of freedom
and can be applied directly to the full PES of Qu and Bowman \cite{Qu2016formic} without further approximation.

Until now theoretical calculations of the splitting have generally varied by at least a factor of 2 
from the observed splittings.
This is due in part to less accurate PESs but also to approximations made to the quantum dynamics.
As it has not been possible to perform an exact quantum calculation for the full-dimensional system,
some compromise has to be made.
These typically include either reduced-dimensionality approaches
\cite{chang1987analysis, %
shida1991reaction, %
vener2001vibrational, %
barnes2008effects, %
matanovic2008generalized, %
malis2009computational,
Luckhaus2006formic,
Luckhaus2010formic, %
jain2015tunneling} %
or approximate semiclassical methods.
\cite{tautermann2004ground,
smedarchina2004calculation,
milnikov2005ground}
Although the ring-polymer instanton approach also employs a semiclassical approximation,
its application differs from previous semiclassical studies
in that the optimal pathway is located on the full PES,
whereas in \Ref{tautermann2004ground} the pathway is defined as an interpolation between the minimum-energy pathway and the sudden approximation,
in \Ref{smedarchina2004calculation} the PES is modelled by a quartic double-well coupled to harmonic modes,
and in \Ref{milnikov2005ground} the pathway is optimized on a less-accurate DFT surface which is only partially corrected for with high-level CCSD(T) calculations along the path.

In order to understand why the previous theoretical approaches have failed
to predict the tunnelling splitting quantitatively,
we make a systematic study of the problem
applying the same instanton approach both to full- and reduced-dimensional systems.
By comparison with exact quantum results where possible,
this analysis can provide an estimate of the errors made both by a reduced-dimensionality approach
or by the instanton approximation itself.
In this way,
we show that the full-dimensional instanton method is expected to give more accurate predictions
than exact calculations in reduced dimensionality unless almost all of the modes are included.
This is confirmed by comparison of the instanton predictions for the tunnelling splittings
with experimental measurements
which agree within our error estimation.

\section{Theory}

The ring-polymer instanton approach \cite{tunnel}
provides a semiclassical approximation to the path-integral description of a tunnelling process.
Rather than enumerating over all possible paths, as in Feynman's exact theory, \cite{Feynman}
one simply locates the dominant pathway
and employs a steepest-descent integration around it.
\cite{ABCofInstantons,Benderskii,Milnikov2001}
For high barriers and small tunnelling splittings, such as are found in this case,
the method is expected to give an excellent approximation to the exact quantum mechanical result.
\cite{tunnel}

The dominant pathway is found, as described in \Ref{tunnel},
by optimizing a fictitious linear polymer of beads
where each bead is a representation of the system and the beads are connected by mass-dependent harmonic springs.
Fluctuations about the pathway are described within a harmonic approximation
by diagonalizing the banded mass-weighted Hessian matrix of the polymer.
One thus treats the anharmonicity along the pathway explicitly but neglects anharmonic effects orthogonal to it.
That this is a double proton transfer causes no difficulties to the instanton method as it is well known to proceed by a concerted mechanism \cite{Ivanov2015formic}
and is therefore treated as for any other double well system.

The reduced-dimensional system is defined as described in Refs. \onlinecite{Kamarchik2009reduced}
and \onlinecite{Qu2016formic}.
This procedure can be summarized as follows:
the internal modes are defined by the normal modes at the saddle point
and their values are computed along the minimum-energy pathway.
The reduced-dimensional PES can then be defined as a function of a set of specified modes,
with the unspecified modes given their value
found by interpolating along the minimum-energy pathway
with respect to the imaginary mode only.
The symmetry of the problem was rigorously preserved to ensure that the two minima are numerically exactly degenerate.

The minima of the formic acid dimer
have $C_\mathrm{2h}$ point-group symmetry and its symmetry elements
will be preserved along the tunnelling pathway.
At the mid-point of the pathway, as at the saddle point,
extra symmetry elements appear and the point group expands to $D_\mathrm{2h}$.
However, 
we use the irreps of the $C_\mathrm{2h}$ point group to label the modes at both the minima and saddle point 
along with an assignment of the major components of the displacement vectors obtained by inspection.
\cite{Luckhaus2010formic}

\begin{table}[h]
\begin{threeparttable}
\small
  \caption{\ Modes used for $f$-dimensional calculations
  following the notation of \Ref{Luckhaus2010formic}
  with irreps, $\Gamma$, of the $C_\mathrm{2h}$ point group.
  Frequencies, $\omega$, and the mode displacements, $Q$,
  are given for \HCOOH\ at the stationary points.
  The $\mathrm{A_g}$ modes are ordered according to the magnitude of $Q_\text{min}$
  and the remaining modes according to the magnitude of $\Delta\omega=\omega_\text{min}-\omega_\text{sad}$.
  }
	\begin{tabular}{d{2.0}lcd{3.1}d{4.0}d{5.0}}
	\hline
	\head{$f$} & mode\tnote{a} & \head{$\Gamma$} & \head{$|Q_\text{min}|$} & \head{$\omega_\text{min}$/\wn} & \head{$\omega_\text{sad}$/\wn} \\
	\hline

1 & $\nu_\mathrm{OH}$ & $\mathrm{A_g}$ & 45.7 & 3232 & -1355\iu \\
2 & $\nu_\mathrm{R}$ & $\mathrm{A_g}$ & 97.3 & 209 & 514 \\
3 & $\beta_\mathrm{R}$ & $\mathrm{A_g}$ & 67.6 & 167 & 219 \\
4 & $\beta_\mathrm{OCO}$ & $\mathrm{A_g}$ & 28.1 & 693 & 774 \\
5 & $\nu_\mathrm{CO}(+)$ & $\mathrm{A_g}$ & 8.2 & 1255 & 1408 \\
6 & $\beta_\mathrm{OH}$ & $\mathrm{A_g}$ & 4.7 & 1481 & 1691 \\
7 & $\nu_\mathrm{CO}(-)$ & $\mathrm{A_g}$ & 4.2 & 1715 & 1749 \\
8 & $\beta_\mathrm{CH}$ & $\mathrm{A_g}$ & 1.2 & 1408 & 1397 \\
9 & $\nu_\mathrm{CH}$ & $\mathrm{A_g}$ & 0.1 & 3095 & 3101 \\
10 & $\nu_\mathrm{OH}$ & $\mathrm{B_u}$ & 0 & 3326 & 1241 \\
11 & $\delta_\mathrm{OH}$ & $\mathrm{A_u}$ & 0 & 970 & 1400 \\
12 & $\delta_\mathrm{OH}$ & $\mathrm{B_g}$ & 0 & 956 & 1341 \\
13 & $\beta_\mathrm{R}$ & $\mathrm{B_u}$ & 0 & 275 & 592 \\
14 & $\beta_\mathrm{OH}$ & $\mathrm{B_u}$ & 0 & 1448 & 1604 \\
15 & $\nu_\mathrm{CO}(+)$ & $\mathrm{B_u}$ & 0 & 1258 & 1404 \\
16 & $\beta_\mathrm{OCO}$ & $\mathrm{B_u}$ & 0 & 716 & 814 \\
17 & $\delta_\mathrm{R}$ & $\mathrm{B_g}$ & 0 & 254 & 317 \\
18 & $\delta_\mathrm{R}$ & $\mathrm{A_u}$ & 0 & 170 & 226 \\
19 & $\nu_\mathrm{CO}(-)$ & $\mathrm{B_u}$ & 0 & 1780 & 1743 \\
20 & $\delta_\mathrm{CH}$ & $\mathrm{A_u}$ & 0 & 1100 & 1079 \\
21 & $\delta_\mathrm{CH}$ & $\mathrm{B_g}$ & 0 & 1084 & 1065 \\
22 & $\beta_\mathrm{CH}$ & $\mathrm{B_u}$ & 0 & 1406 & 1395 \\
23 & $\tau_\mathrm{R}$ & $\mathrm{A_u}$ & 0 & 70 & 80 \\
24 & $\nu_\mathrm{CH}$ & $\mathrm{B_u}$ & 0 & 3097 & 3106 \\

	\hline
	\end{tabular}
	\begin{tablenotes}
	\item[a] $\nu$ is stretch, $\beta$ is in-plane bend, $\delta$ is out-of-plane bend, $\tau$ is torsion, R is intermolecular, $\pm$ is symmetric or antisymmetric.
	\label{tab:modes}
	\end{tablenotes}
\end{threeparttable}
\end{table}

We will consider a set of reduced-dimensional systems each with a different numbers of modes, $f$.
The order in which modes are included is specified in \tref{modes};
for graphical representations of these see the supplementary material of \Ref{Qu2016formic}.
This list has been compiled 
with the intention that the modes with the most significant contributions to the tunnelling splitting appear towards the top,
according to the scheme outlined below.

Only the $\mathrm{A_g}$ modes are coupled to the proton transfer mode
and thus only they will be excited along the instanton pathway.
These first 9 modes are ordered according to the magnitude of their displacement at the minimum-energy configurations.
All other modes remain at 0 along the pathway.
However, they may still effect the tunnelling splitting by changing the effective zero-point energy along the path.
We order these modes according to the magnitude of the difference in their vibrational frequencies between the minimum and saddle point, $\Delta\omega=\omega_\text{min}-\omega_\text{sad}$.
The order of the first few modes is the same as was chosen in
\Ref{Qu2016formic} %
which makes comparison of results with this study possible.

Note that in \tref{modes} we have ordered the modes according to their properties in the \HCOOH\ isotopomer.
After isotopic substitution, the frequencies and displacements at equilibrium change somewhat which
means that the modes no longer strictly follow the order specified by the magnitudes of $Q_\text{min}$ and $\Delta\omega$.
However, this does not make significant changes to the conclusions
and we have chosen to keep the order the same for all isotopomers to make comparison and presentation of the results simpler.

\section{Results}

We ran two types of instanton calculations,
one in Cartesian coordinates and the other using the reduced-dimensionality modes.
The only difference in the algorithm is that the 6 frequencies
related to translations and rotations of the linear polymer only need to be excluded
in the former case.
In both cases, 
a length of $\beta\hbar=20000\,\mathrm{a.u.}$ with $2048$ beads was found to give converged results to two significant figures.

Because it is known by symmetry prescriptions
that modes with irreps other than $\mathrm{A_g}$ will not be excited along the pathway,
the instanton optimized with nine degrees of freedom
can be used for all calculations with $f \ge 9$.
After computing the full Hessians along this pathway,
careful data analysis made it possible to obtain the $f={9,\dots,24}$ splittings
without extra computation.

\begin{figure}
\centering
  \includegraphics[width=8.3cm]{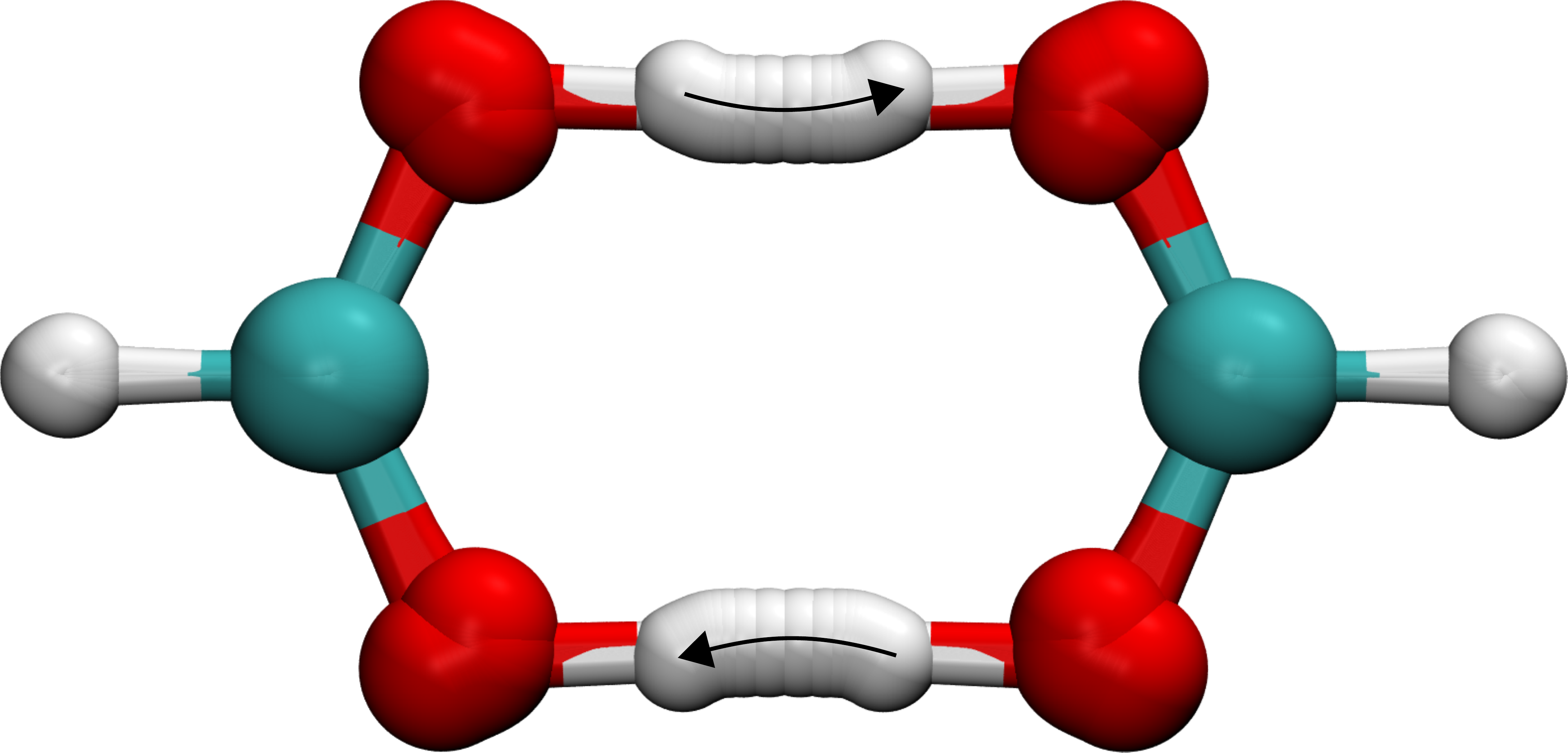}
  \caption{Representation of the \HCOOH\ instanton pathway
  using conflated snapshots of the configurations.
  It is clear that a certain amount of movement of the skeletal atoms is coupled to the proton transfer.
  }
  \label{fig:slug}
\end{figure}

\begin{figure}
\centering
  \includegraphics{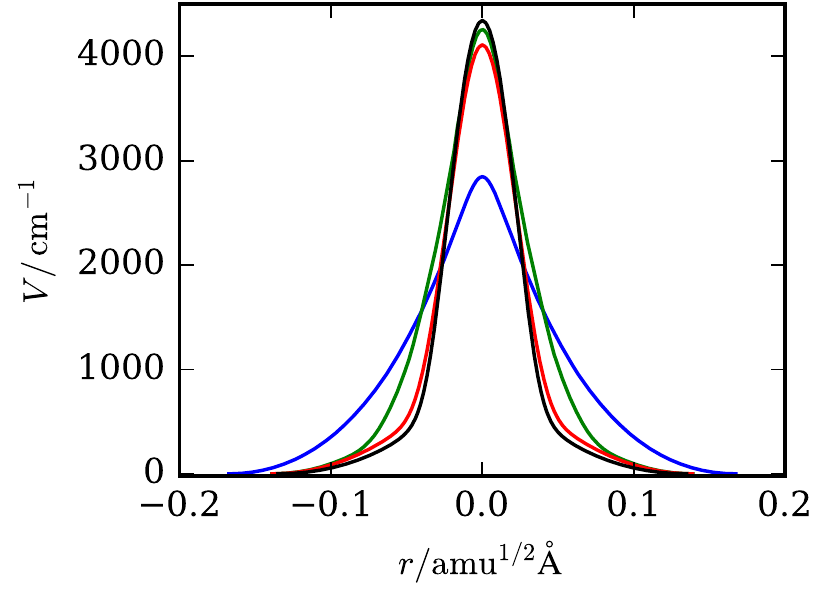}
  \caption{Plot of the potential energy, $V$,
  against the mass-weighted distance, $r$, 
  along the instanton paths computed for \HCOOH\
  with different numbers of modes:
  $f=1$ (blue), $f=2$ (green), $f=3$ (red) and \mbox{$f\ge9$} (black).
  As the higher modes do not affect the instanton pathway the last curve is the same as the full-dimensional case.}
  \label{fig:pot}
\end{figure}

In \fig{slug} we show a graphical representation of the instanton for \HCOOH\
and in \fig{pot} the potential energy along the pathway.
In \fig{pot}, one can see that the first three modes are able to
characterize the full-dimensional instanton pathway fairly well.
It is noted that the instanton does not pass through the saddle point
but rather chooses to cut the corner to reduce the width of the barrier
at the expense of increasing the barrier height.
This occurs because the action for the latter case is lower and the
ring-polymer instanton method automatically finds the minimum action pathway.
Note that separate instanton optimizations are required for the other isotopomers,
which give similar graphics,
the major difference being that for \DCOOD\ and \HCOOD\ the \mbox{$f\ge9$} effective barrier is slightly wider and about 500\,\wn\ lower.

\begin{table}
\begin{threeparttable}
\small
  \caption{\ Ground-state tunnelling splittings calculated by the instanton method in cm$^{-1}$.  In parentheses are results from reduced-dimensionality quantum mechanics or experiments as indicated.}
  \label{tab:results}
  \begin{tabular*}{0.5\textwidth}{@{\extracolsep{\fill}}lcccc}
    \hline
	 $f$ & \HCOOH & \DCOOH & \HCOOD & \DCOOD \\
    \hline
	 1 & \double{0.47}{(0.44)\tnote{a}} & \double{0.44}{(0.41)\tnote{a}} & \double{0.017}{(0.017)\tnote{b}} & \double{0.016}{(0.015)\tnote{a}} \\
	 2 & \double{0.17}{(0.16)\tnote{a}} & \double{0.19}{(0.15)\tnote{a}} & \double{0.0052}{(0.0043)\tnote{b}} & \double{0.0040}{(0.0038)\tnote{b}} \\
	 3 & \double{0.037}{(0.032)\tnote{a}} & \double{0.025}{(0.028)\tnote{a}} & 0.00060 & \double{0.00058}{($\sim$0.0003)\tnote{a}} \\
	 4 & \double{0.047}{(0.037)\tnote{a}} & 0.030 & 0.00060 & 0.00057 \\
	 5 & 0.048 & 0.029 & 0.00069 & 0.00066 \\
	 6 & 0.031 & 0.023 & 0.00046 & 0.00042 \\
	 7 & 0.019 & 0.016 & 0.00034 & 0.00031 \\
	 8 & 0.018 & 0.017 & 0.00033 & 0.00031 \\
9 & 0.018 & 0.017 & 0.00033 & 0.00031 \\
10 & 0.28 & 0.24 & 0.0032 & 0.0030 \\
11 & 0.12 & 0.099 & 0.0017 & 0.0014 \\
12 & 0.055 & 0.045 & 0.00085 & 0.00069 \\
13 & 0.033 & 0.028 & 0.00047 & 0.00041 \\
14 & 0.024 & 0.020 & 0.00032 & 0.00029 \\
15 & 0.033 & 0.031 & 0.00036 & 0.00033 \\
16 & 0.030 & 0.029 & 0.00047 & 0.00045 \\
17 & 0.026 & 0.025 & 0.00041 & 0.00039 \\
18 & 0.023 & 0.023 & 0.00035 & 0.00034 \\
19 & 0.016 & 0.016 & 0.00027 & 0.00026 \\
20 & 0.015 & 0.015 & 0.00022 & 0.00025 \\
21 & 0.016 & 0.015 & 0.00022 & 0.00025 \\
22 & 0.014 & 0.014 & 0.00021 & 0.00021 \\
23 & 0.014 & 0.014 & 0.00022 & 0.00022 \\
24 & 0.013 & 0.014 & 0.00020 & 0.00020 \\
	 \text{full} & \double{0.014}{(0.016)\tnote{c}} & \double{0.014}{(0.0125)\tnote{d}} & 0.00021 & \double{0.00021}{($<$0.002)\tnote{e}} \\
    \hline
  \end{tabular*}
\begin{tablenotes}
  \item[a] quantum dynamics from \Ref{Qu2016formic}
  \item[b] quantum dynamics from this work
  \item[c] experimental Refs. \onlinecite{Ortlieb2007HCOOH} and \onlinecite{Goroya2014HCOOH}
  \item[d] experimental Ref. \onlinecite{Madeja2002DCOOH} reassigned by Ref. \onlinecite{Ortlieb2007HCOOH}
  \item[e] experimental \Ref{Gutberlet2008DCOOD}
	\end{tablenotes}
\end{threeparttable}
\end{table}

Table~\ref{tab:results} presents the results from the instanton calculations for all the systems studied.
It is seen that the instanton results agree within about 20 percent of the 
reduced-dimensionality quantum calculations where they are available.
This gives us an estimate of the error caused by the instanton approximation which can be assumed to be approximately the same for larger values of $f$ as well.
In the case of \HCOOD\, which was not treated in that study, we ran our own one- and two-dimensional DVR (discrete-variable representation) \cite{Light1985DVR} calculations
and also report a converged $f=2$ result for \DCOOD. %

Even though modes $f\ge10$ are forbidden by symmetry rules to affect the instanton pathway itself,
including some of them
makes an enormous change to the predicted tunnelling splitting.
Three modes which have the largest effect are the 10th, 11th and 12th degrees of freedom.
This is because their modes excite motion of the central protons orthogonal to the pathway
and will feel different force constants at different points along the instanton.

It so happens that in this example, this effect is mostly compensated for by the remaining modes,
but this was not obvious \emph{a priori} and may not be the case in other systems.
This means that it will not be possible to
choose a small set of modes to define a reduced-dimensionality model
which can be used to obtain an accurate result.

It is clear from the magnitude of $\Delta\omega$
that including the 10th mode, for instance, will dramatically increase the splitting.
However, $\Delta\omega$ is not a completely reliable estimator of the effect of adding extra modes
and would incorrectly predict that the 15th mode would decrease the splitting.
This is because the instanton does not actually pass through the saddle point due to corner cutting,
and it is thus difficult to quantitatively predict the effect that adding extra degrees of freedom will have on the splitting
without performing the instanton calculation.

There is a subtle change in the predicted splitting on going from 24 dimensions to
the full 30 because one of the rotational modes has $\mathrm{A_g}$ symmetry and couples to the pathway.
However, this effect is relatively minor as the
rotational mode affects the central hydrogens much less than the heavier skeletal atoms which remain approximately fixed,
and the moment of inertia hardly changes along the pathway.\cite{tunnel,water}

The full-dimensional results are in good agreement with measurements,
where the experimental splittings are known.
It is not surprising that there is only a small effect on isotopically substituting the outer hydrogens
because the motion of this atom is far from the proton transfer and thus uncoupled.
Therefore, as expected, the action only increases slightly (by $0.07\,\hbar$ in both cases) on deuteration.
This is in agreement with the reassignment
of the experimentally determined \DCOOH\ splitting \cite{Madeja2002DCOOH}
from 0.0029\,\wn\ to 
0.0125\,\wn\ as suggested in \Ref{Ortlieb2007HCOOH}.

However, in contrast to the observations, the calculations predict a
near equivalence between the tunnelling splittings in
\HCOOH\ and \DCOOH\ or between \HCOOD\ and \DCOOD.
This was unexpected as the instanton approach has previously been found to be particularly good at accurately predicting the effect of isotopic substitution
\cite{hexamerprism}
and any errors in the PES are expected to cancel out in these cases.
The slight error which causes this coincidence
must generate from the neglect of anharmonicity
orthogonal to the instanton pathway
which differs depending on the mass of the external hydrogen.

On the other hand, substituting the central hydrogens affects the tunnelling pathway itself,
which is treated fully anharmonically,
and considerably increases the action (by $3.40\,\hbar$ in both cases).
This effect is well described by the instanton approach
as was also seen in previous studies. \cite{water}

To remove the approximation of the anharmonicity without resorting to reduced-dimensionality approaches,
it would be necessary to perform either a path-integral \cite{Matyus2016tunnel1,Matyus2016tunnel2}
or DMC computation.
This would however be a significantly longer computation,
especially in the case of DMC,
which due to the small tunnelling splitting, would require excellent convergence of the statistics.

For a rough analysis of the errors caused by the PES,
we note that the best ab initio barrier height available \cite{Ivanov2015formic} is 2903\,\wn\ 
whereas that predicted by the fitted surface is 2848\,\wn. \cite{Qu2016formic}
Scaling the whole potential-energy surface by 
a factor of $\alpha=2903/2848\approx1.02$
would only change the \HCOOH\ instanton action, \mbox{$S=\int \sqrt{2V}\rmd r$}, from $14.87\,\hbar$ to $15.01\,\hbar$
and hence reduce the tunnelling splitting,
which is proportional to $\eu{-S/\hbar}$, by about 10 percent in this case.
To avoid the error of fitting the PES,
one could in principle also perform the ring-polymer instanton calculation
with on-the-fly calculation of the electronic structure.
The necessary algorithms are available in Molpro \cite{Molpro,Molpro-WIRES,HCH4}
but would require significantly more computational effort than using the fitted PES\@.
Recent improvements in the efficiency \cite{Cvitas2016instanton} of the ring-polymer instanton method may make this a viable method for future studies on systems without fitted PESs.

We are thus able to identify
the cause of the discrepancy between the 
3 or 4-dimensional quantum calculations of Qu and Bowman \cite{Qu2016formic}
and the experimental observations.
It is due in the main to the reduced-dimensionality treatment
rather than with any errors in the fitted PES\@.
Like their calculations,
the instanton approach also predicts a splitting a factor of 2 too large when only 3 or 4 modes are included.
However, by increasing the number of modes to full dimensionality
we are able to significantly reduce the error
and obtain results much closer to those found in experimental studies.

\section{Conclusions}

For the formic acid dimer, %
the ring-polymer instanton method has been used to predict ground-state tunnelling splittings with less than about 20 percent error when compared to exact approaches
on reduced-dimensionality systems.
However, the instanton approach is also able to treat the system in full-dimensionality,
optimize the tunnelling pathway without \emph{a priori} constraints
and has calculated tunnelling splittings close to the experimentally-measured values.
We thus confirm the reassignment of the \DCOOH\ splitting to be the larger of the two possible values
suggested by \Ref{Madeja2002DCOOH}.
As was later pointed out, 
to a very good approximation, the tunnelling splitting should
be independent upon deuteration of the outer H atom, \cite{Ortlieb2007HCOOH}
and this is in agreement with our calculations.

Our study concludes that the majority of modes will have a non-negligible effect
on the tunnelling splitting which makes it extremely difficult to obtain high-accuracy
predictions by increasing the dimensionality towards convergence.
It thus seems that a reduced-dimensionality approach,
even one employing an exact quantum dynamics method,
will not be able to obtain results more accurately than the instanton
unless almost all the degrees of freedom are included.

From an analysis of the errors inherent in the PES,
we expect it to be the cause of only about 10 percent error,
whereas the instanton approximation makes about a 20 percent error
as is clearly seen when comparing with the reduced-dimensionality quantum results where available.
We thus conclude that the observed 20 percent difference between the calculated full-dimensional splittings
and those measured experimentally is due to a combination of these well-understood errors
and that these instanton calculations represent the most accurate theoretical treatment performed on this system to date.
Previous theoretical treatments have used less accurate PESs and either introduced errors by using reduced-dimensionality approaches
or by approximating the tunnelling pathway.

Although the \DCOOD\ splittings were beyond the experimental resolution of \Ref{Gutberlet2008DCOOD},
they may be observable to future experiments.
We expect our prediction for this isotopomer to be as accurate as for the others
and can thus be taken as a target resolution for an experimental setup.

\section{Acknowledgements}
The author would like to thank
Chen Qu and Joel Bowman for useful discussions.

\end{document}